\documentclass[aps, prd, twocolumn, superscriptaddress, nofootinbib, amsmath, amssymb]{revtex4-1}

\usepackage{graphicx}  
\usepackage{mathrsfs}
\usepackage{bbold} 
\usepackage{braket}
\usepackage[normalem]{ulem} 
\usepackage[breaklinks=true,colorlinks,citecolor=blue,linkcolor=blue,urlcolor=blue]{hyperref}

\newcommand{\ep}{\varepsilon}
\newcommand{\Id}{\mathbb{1}}

\begin{document}

\title{Quantum computation of thermal averages in the presence of a sign problem}

\author{Giuseppe~Clemente}
\affiliation{Dipartimento di Fisica dell'Universit\`a di Pisa and INFN
  - Sezione di Pisa, Largo Pontecorvo 3, I-56127 Pisa, Italy.}

\author{Marco~Cardinali}
\affiliation{Dipartimento di Fisica dell'Universit\`a di Pisa and INFN
  - Sezione di Pisa, Largo Pontecorvo 3, I-56127 Pisa, Italy.}

\author{Claudio~Bonati}
\affiliation{Dipartimento di Fisica dell'Universit\`a di Pisa and INFN
  - Sezione di Pisa, Largo Pontecorvo 3, I-56127 Pisa, Italy.}

\author{Enrico Calore}
\affiliation{Universit\`a degli Studi di Ferrara and INFN - Sezione di Ferrara,
Via Saragat 1, I-44122 Ferrara, Italy}

\author{Leonardo~Cosmai}
\affiliation{INFN - Sezione di Bari, I-70126 Bari, Italy}

\author{Massimo~D'Elia}
\affiliation{Dipartimento di Fisica dell'Universit\`a di Pisa and INFN
  - Sezione di Pisa, Largo Pontecorvo 3, I-56127 Pisa, Italy.}

\author{Alessandro~Gabbana}
\affiliation{Universit\`a degli Studi di Ferrara and INFN - Sezione di Ferrara,
Via Saragat 1, I-44122 Ferrara, Italy}

\author{Davide~Rossini}
\affiliation{Dipartimento di Fisica dell'Universit\`a di Pisa and INFN
  - Sezione di Pisa, Largo Pontecorvo 3, I-56127 Pisa, Italy.}

\author{Fabio~Sebastiano~Schifano}
\affiliation{Universit\`a degli Studi di Ferrara and INFN - Sezione di Ferrara,
Via Saragat 1, I-44122 Ferrara, Italy}

\author{Raffaele~Tripiccione}
\affiliation{Universit\`a degli Studi di Ferrara and INFN - Sezione di Ferrara,
Via Saragat 1, I-44122 Ferrara, Italy}

\author{Davide~Vadacchino}
\affiliation{Dipartimento di Fisica dell'Universit\`a di Pisa and INFN
  - Sezione di Pisa, Largo Pontecorvo 3, I-56127 Pisa, Italy.}

\collaboration{QuBiPF Collaboration}
\thanks{giuseppe.clemente@pi.infn.it \\
marco.cardinali@pi.infn.it \\
claudio.bonati@unipi.it \\
enrico.calore@fe.infn.it \\
leonardo.cosmai@ba.infn.it \\
massimo.delia@unipi.it \\
gbblsn@unife.it \\
davide.rossini@unipi.it \\
schifano@fe.infn.it \\
tripiccione@fe.infn.it \\
davide.vadacchino@pi.infn.it}

\begin{abstract}
  We illustrate the application of Quantum Computing techniques 
  to the investigation of the thermodynamical properties
  of a simple system, made up of three quantum spins with frustrated 
  pair interactions and affected by a hard sign problem when treated
  within classical computational schemes. We show how quantum algorithms
  completely solve the problem, and discuss how this can apply to more
  complex systems of physical interest, with emphasis on the possible
  systematics and on their control.
\end{abstract}

\maketitle

\section{Introduction}

The computation of thermal averages for a quantum mechanical system is a common
challenge to many different fields of modern physics, in diverse areas such as condensed
matter or fundamental interactions.  In most cases, the issue is connected to
the determination of the phase diagram of the system: this is the case, for
instance, of Quantum Chromodynamics (QCD), the theory which describes strong
interactions within the standard model of particle physics. In this context,  the QCD phase diagram is related to fundamental questions regarding the properties of the Universe in its early stages, or the internal structure and composition of compact astrophysical objects, such as neutron stars~\cite{Shapiro1983,Rajagopal:2000wf}.

In the absence of an exact solution, the strongly interacting nature of the
system generally makes analytical tools, such as perturbation theory,
unfeasible, and demands for a numerical approach. Monte-Carlo (MC)
algorithms represent a possible solution to the problem: in a nutshell, the
computation of quantum thermal averages is brought to the form of a classical
computation.  Namely, after properly choosing the basis of the Hilbert space
(the so-called computational basis), one performs a classical computation by
means of classical MC methods, when possible.  The path-integral formulation of
quantum mechanics provides a suitable framework where this can be done
systematically.  The partition function of the quantum thermal system is
rewritten in terms of an Euclidean path integral, which resembles the
partition function of a classical statistical system in one more dimension
(Euclidean time); Euclidean paths, or field configurations, are
assigned a Boltzmann-like weight, corresponding to the exponential of minus the
action. The path integral, in particular thermal averages, are then computed
using standard tools, such as MC importance sampling. 

Unfortunately, the prescription depicted above fails dramatically when the
weight turns out to be non-positive defined or, even worse, complex. This
constitutes the infamous {\it sign problem}, which prevents progress in many
fields, such as the accurate determination of the QCD phase diagram at finite
baryon density \cite{Philipsen:2010gj, Aarts:2015tyj, Ding:2015ona}, and can be
shown to be a NP-hard task~\cite{Troyer:2004ge}. The problem is usually related
to the path-integral formulation itself, or, in other words, to the choice of
the computational basis. Indeed, in the canonical formulation of the partition
function, all the Hamiltonian eigenstates have a positive weight, as it stems
from the hermiticity of the Hamiltonian itself, or of other possible conserved
charges coupled to chemical potentials. However, determining the spectrum and
the eigenstates of the Hamiltonian usually implies a similar or even greater
level of computational difficulty.

The sign problem is often taken as one of the compelling motivations for the
development of Quantum Computing (QC).  Moving forward from the original
proposal of Feynman~\cite{Feynman1982}, QC has been
nowadays exploiting the possibility to operate algorithms based on quantum
logic gates, aimed at either solving certain tasks more efficiently than
classically, or in general reproducing the real-time evolution of a system on a
digital, or analog, platform~\cite{NielsenChuang, BookQC}.  Cornerstone
examples are the Lloyd's proposal for the quantum simulation of the
Schr\"odinger equation~\cite{lloyd}, the Shor's algorithm for integer
factorization~\cite{shor}, and the Grover's algorithm for searching from an
unstructured database~\cite{grover}.  Our current investigation revolves around
the question on how one can make use of a quantum computer in order to
reproduce the quantum thermal averages and then explore the phase diagram,
i.e., on how the sign problem can be solved in practice.

The reconstruction of quantum thermal averages via a quantum computer has been
object of a rather intense theoretical activity in the past few
years~\cite{DiVincenzo, Poulin, Boixo, QMS_paper, Aspuru-Guzik, Eisert, Cirac,
Brandao, Neven, Somma, Freericks, Anschuetz_Cao, Moussa}. Here we focus on a
specific implementation, proposed in Ref.~\cite{QMS_paper}, of the so-called
Quantum Metropolis Sampling (QMS), where the main idea is to exploit quantum
parallelism in order to sample directly in the eigenbasis of the Hamiltonian.
The purpose of this study is to implement the algorithm for a prototype system
where one finds a sign problem in the path-integral formulation, showing that
this approach completely solves the problem and investigating the possible
systematics of the method. In particular, we consider a system of three quantum
spins coupled to each other via antiferromagnetic interactions~\cite{Wannier,
Balents}, a model that has already been used as a QC testbed in several works,
and in particular in quantum simulators using trapped ions \cite{ExpTI1,
ExpTI2} and nuclear magnetic resonance~\cite{ExpNMR}. 

While it is almost trivial to diagonalize the system Hamiltonian and perform a
classical sign problem-free computation, the way in which the quantum computer
solves it is general and applies, at least in principle, to quantum systems of
arbitrary complexity.  Indeed, as we will better explain in the following,
everything relies on the possibility of performing the quantum phase estimation
(QPE) algorithm for the system at hand: it is at this point that the quantum
advantage becomes manifest, since the algorithm is able to do that with no need
for an explicit diagonalization of the Hamiltonian.

This paper is organized as follows. In Sec.~\ref{description} we give a brief
description of the model, showing how a discrete path integral (Trotter-Suzuki
decomposition~\cite{Trotter1959,Suzuki1976}) performed in the standard
computational basis leads to a sign problem.  Section~\ref{qmsalgo} provides a
short summary of the QMS algorithm adopted in this study.
Section~\ref{results} contains a summary of the results obtained with our
model, while the possible systematics related to the algorithm are discussed in
Sec.~\ref{systematics}.  We conclude with Sec.~\ref{conclusions}, summarizing
our results and discussing possible extensions to more complex systems.

The quantum algorithm illustrated in this study has been implemented on a
quantum simulator (Simulator for Universal Quantum Algorithms, SUQA) developed
on purpose by one of the authors (GC), which is inspired to the well known
Qiskit simulator\footnote{We used our own simulator to add some ad-hoc features
regarding the implementation of hybrid quantum-classical operations.}
\cite{qiskit}.

\section{Description of the Model}
\label{description}

In this section we introduce the system to be used as a playground for testing
the QMS algorithm. It is made up of three quantum spin-1/2 variables, with
frustrated pair interactions. Its Hamiltonian reads
\begin{equation}
  H = J ( \sigma_x \otimes \sigma_x \otimes \Id + \sigma_x 
  \otimes \Id \otimes \sigma_x  +  \Id \otimes \sigma_x \otimes \sigma_x )\, ,
  \label{eq:HamSpinSys}
\end{equation}
where $\Id$ stands for the $2 \times 2$ identity operator, $\sigma_j$ (with $j
= x,y,z$) denotes the usual spin-1/2 Pauli matrices, and $J$ is a positive
parameter (in order to make the couplings antiferromagnetic).  Each of the
$\Id$ or $\sigma_x$ operators acts on the spin corresponding to its position in
the tensor product, and the full Hilbert space of the system is spanned by eight basis states.
A possible choice, which makes the problem trivial, is to select as a
basis the eight products of eigenstates (for each spin) of the $\sigma_x$
operator: this is also a basis of eigenstates of the Hamiltonian, which has
only two energy levels, the lower one with energy $E_0 = -J$ and degeneracy 6
(two spins aligned in either direction and opposite to the third spin), the
higher one with energy $E_1 = 3 J$ and degeneracy 2 (three spins aligned in either
direction).

We are however interested in working with the standard computational basis,
made up of eigenstates of the $\sigma_z$ operators for all spins,
which is the one where the sign problem appears.  In the following we will
label, for simplicity, each of the eight states of this basis by its decimal
correspondent, i.e.~starting from $\ket{\mathbf{0}} \equiv \ket{000}$ to
$\ket{\mathbf{7}} \equiv  \ket{111}$. In this basis, the partition function of
the system reads
\begin{equation}
  Z = \mathrm{Tr} \big[ \mathrm{e}^{-\beta H} \big]
  = \sum _{{\bf \alpha = 0}}^{\bf 7} \bra{ {\bf \alpha}}\mathrm{e}^{-\beta H}\ket{ {\bf \alpha}} \, ,
\end{equation}
where $\beta$ is the inverse temperature (we adopt natural units, $\hbar=k_B=1$).
The path-integral approach consists in rewriting the expression above by
partitioning the exponential into the product of $N$ smaller ``evolution
steps'' (Trotter-Suzuki decomposition~\cite{Trotter1959,Suzuki1976}),
and inserting $N-1$ projectors over a complete set of states between each couple
of evolution steps. Doing so, we arrive to the well known expression
\begin{equation}
  Z = \sum_{(\{\alpha_i \})}\prod_{i=1}^{N}\bra{\alpha_{i+1}}\mathrm{e}^{-\frac{\beta H}{N}}\ket{\alpha _{i}}\, ,
  \label{eq:Z_ham}
\end{equation}
where $\alpha_{N+1} \equiv \alpha_{1}$ and $(\{ \alpha_i \})$ is the space of
all possible configurations, which we can interpret as periodic paths labelled
by a discrete Euclidean time variable $i$. If the product of matrix elements
turns out to be positive for each path, then the sum over all possible
Euclidean paths appearing in Eq.~(\ref{eq:Z_ham}) can be sampled by a classical
MC algorithm.  However, it is easy to show that this is not the case for this
system, i.e.~that it is always possible to find paths for which the product of
matrix elements turns out to be negative. 

In order to show this fact, let us rewrite the elementary evolution
step in the following form:
\begin{equation}
  \mathrm{e}^{-\frac{\beta H}{N}} = 
  \frac{1}{4}\left[\left(e^{-3\, \ep} + 3\, e^{\ep} \right) \Id
    + \left(e^{-3\,\ep} - e^{\ep} \right)\frac{H}{J} \right]\, ,
  \label{eq:rewrite}
\end{equation}
where $\ep \equiv \beta J / N$ [details about the derivation of
Eq.~(\ref{eq:rewrite}) are reported in Appendix~\ref{app:A}].  Since the
Hamiltonian is composed of a symmetric combination of pair-products of
$\sigma_x$ operators, which flip the $\sigma_z$-component of the corresponding
spin, it should be clear that the only non-zero non-diagonal matrix elements of
$\exp(-{\beta H}/{N})$ are those between states differing by exactly two spin
flips. This fact enables to split the computational basis into two disjoint
sets:
\begin{subequations}
\begin{align}
  {\textbf{Set\ 1}\ :\ \ } &
  \{ \ket{\mathbf{0}}, \ket{\mathbf{3}}, \ket{\mathbf{5}}, \ket{\mathbf{6}} \}\, . \\
  {\mathbf{Set\ 2}\ :\ \ } &
  \{ \ket{\mathbf{1}}, \ket{\mathbf{2}}, \ket{\mathbf{4}}, \ket{\mathbf{7}} \}\, .
\end{align}
\end{subequations}

Non-diagonal elements within the same set are all equal and negative, and in
particular their common value is $-e^\ep ( 1 - e^{-4 \ep} )/4$, while
non-diagonal elements across the two sets are exactly zero.  In contrast,
diagonal elements are all equal to $e^\ep (3 + e^{-4 \ep})/4$.

To summarize, the discrete path-integral sum in Eq.~(\ref{eq:Z_ham}) simplifies
by decoupling into two distinct and simpler sums, where paths run either within
$\textbf{Set\ 1}$ or within $\textbf{Set\ 2}$. However, within each sum, paths
presenting an odd number\footnote{It is always possible to find such paths,
since one can also insert an arbitrary number of diagonal matrix elements along
the Euclidean time.} of flips between different states carry a negative weight.
The contribution of these paths cannot be rewritten in terms of a positive
Boltzmann weight, i.e.~in the form $\sum \exp(-S[\alpha_i])$ where
$S[\alpha_i]$ is a real Euclidean action.  This invalidates the possibility of
using classical methods which approximate the sum by means of a MC sampling of
its terms.

In the QMS algorithm described in the next section, the focus is brought again
to the basis of eigenstates of the Hamiltonian, which are sampled according to
their thermal weight. However this is not done by exploiting the exact solution
of the problem (that, in this case, would be trivial), but in a way which can be
easily generalized to more complex cases. As we shall see below, the key
ingredient to achieve this task will be the QPE
algorithm~\cite{phase_estimation, PhaseEstim}.

\section{The Quantum Metropolis Sampling algorithm}
\label{qmsalgo}

This section is a brief, yet self-contained, overview of the
main steps of the QMS algorithm, to be operated on a quantum computer.
Further details can be found in the original publication, Ref.~\cite{QMS_paper}.

The key ingredient of QC, which potentially enables an exponential
speedup over any known classical algorithm analogue, is quantum parallelism.
Namely, the possibility, for a set of $N$ two-level quantum systems
(the so-called qubits, quantum analogues of classical bits),
to be in a superposition of states, thus simultaneously 
acting on an exponential number $2^N$ of classical configurations.
Such an ensemble of qubits can be suitably manipulated through
a sequence of invertible single-qubit and two-qubit controlled-NOT (CNOT)
quantum logic gates (all mathematically representable as unitary operators);
this constitutes a universal set of quantum gates~\cite{NielsenChuang, BookQC}.

The QMS algorithm is a procedure which enables to perform a quantum MC sampling
of thermal averages, using a sequence of invertible quantum logic gates
on a register of qubits and classical measures on appropriate subportions of it.
More specifically, QMS makes use of QPE in order to record the eigenvalues
of a given quantum Hamiltonian without explicitly knowing the detailed
structure of its eigenvectors.
Before entering the overview of QMS, we discuss the working principle of QPE,
lying at the heart of it.

\subsection{Quantum phase estimation}
\label{sec:QPE}

The QPE algorithm applies to the following general problem.
Suppose a given unitary operator $U$ has an eigenvector $\ket{u}$
with eigenvalue $e^{i \theta}$ ($0 \leq \theta \le 2 \pi$). Also assume that
one is able to prepare the state $\ket{u}$ and to perform
controlled-$U^{2^j}$ gates ($j \in \mathbb{N}$).
The latter operation needs a set of two quantum registers,
the first one (1) containing the $n$ qubits necessary to store $\ket{u}$;
the second one (2) containing an ancillary qubit, such that
\begin{equation}
  \frac{\ket{0}_2 + \ket{1}_2}{\sqrt{2}} \otimes \ket{u}_1
  \xrightarrow{{\rm c}\mbox{-}U^{2^j}} \frac{\ket{0}_2 + e^{i 2^j \theta} \ket{1}_2}{\sqrt{2}} \otimes \ket{u}_1 .
\end{equation}
The purpose of QPE is to obtain the best $l$-bit estimate of $\theta$,
for a given integer value of $l$~\cite{phase_estimation, PhaseEstim}.

To achieve this goal, one discretizes the possible values of $\theta$
choosing a suitable encoding in binary notation with $r$ bits
[up to $O(2^{-(r+1)})$ accuracy].
A sequence of the above controlled-$U^{2^j}$ gates needs to be applied
(varying $j = 0, 1, 2, \ldots$) on $r$ different ancillary qubits,
here referred to as the control register.
Since the quantum register that stores $\ket{u}$ is prepared in an eigenstate
of the operators $U, U^2, U^4, \cdots$, the state of this register never changes,
while the phase factors $e^{i\theta}, \, e^{2i\theta}, \, e^{4i\theta} , \ldots$
are propagated backwards in the control register.
Finally, a quantum Fourier transform on the control register 
can be shown to provide an approximate binary encoding of $\theta$,
where each digit is represented by the state of a specific qubit state of the register.
In particular, the algorithm outputs the best $l$-bit estimate of $\theta$
with probability $> 1-\epsilon$,
if the control register contains $r = l + O(\log(1/\epsilon))$ qubits
and the obtained result is rounded off to its most significant $l$ bits~\cite{phase_estimation}.
For further details on the QPE algorithm, see, e.g., Ref.~\cite{BookQC}.

In the specific implementation of QMS, the QPE procedure is applied to the
unitary operator $U = e^{-iHt}$, which, for a large class of physically relevant
Hamiltonians, can be efficiently simulated on a quantum computer~\cite{lloyd},
after an appropriate discretization of the real-time evolution (Trotterization)~\cite{Trotter1959,Suzuki1976}.
Since an eigenvector $\ket{\psi_\alpha}$ of $H$ with eigenvalue $E_\alpha$
is also an eigenvector of $U(t)$ with eigenvalue $e^{-i E_\alpha t}$,
it is thus possible to encode, in the control register, the eigenvalues
of $H$ without explicitly knowing the detailed structure
of its eigenvectors~\cite{PhaseEstim}.

\subsection{Implementation of Quantum Metropolis Sampling}
\label{sec:QMS}

We consider a generic quantum system that can be represented (or approximated)
by a finite number $n$ of qubits and is described by the Hamiltonian $H$.  In
principle, the eigenvalues of $H$ could take any real value, but for the sake
of simplicity let us assume that they can be exactly represented by a certain
number $r$ of qubits. This is always possible for commensurate energy spacings,
by discretizing their possible values and using the binary representation of
the integers $0$ through $2^r -1$.  As emerging from the discussion above in
Sec.~\ref{sec:QPE}, and will be clarified later in Sec.~\ref{sec:Syst-E},
for systems with energy levels that are not multiple of a certain energy unit
this introduces a systematic error, which however can be made arbitrarily small
by increasing $r$ (at the cost of increasing the complexity of the algorithm).

The idea of the QMS algorithm is to adapt to the QC framework the classical
Markov Chain MC method \cite{Metropolis:1953am}, by generating a sequence of
eigenstates of $H$ distributed according to the Boltzmann statistical weight.
While some steps of the original classical algorithm can be extended to the
quantum case straightforwardly, the accept-reject step requires 
more care since,
due to the no-cloning theorem~\cite{nocloning1, nocloning2}, it is not possible
to have a backup copy of the original configuration to be used if
the proposed update is rejected.

To overcome this difficulty, the QMS algorithm uses four quantum registers: a
register representing the system wavefunction $\ket{\psi}$ with $n$ qubits, two
$r$-qubits registers holding information about the energy before ($\ket{E^{\rm
old}}$) and after ($\ket{E^{\rm new}}$) a proposed update, and a single-qubit
register with information about the acceptance ($\ket{\rm acc}$). All these
registers can be arranged as
\begin{equation}
  \ket{{\rm acc}}_4 \otimes \ket{E^{\rm new}}_3 \otimes \ket{E^{\rm old}}_2 \otimes \ket{\psi}_1 \quad ;
  \label{registers}
\end{equation}
in the following, when omitted, the subscripts corresponding to the various
registers are intended as in~\eqref{registers}. The QMS algorithm then proceeds as
follows:

{\it Step 0:}
At the beginning, the input state~\eqref{registers} has to be initialized to
zero for all registers, except the one holding the system state, which must be
set to an eigenstate $\ket{\psi_k}$ of the Hamiltonian: $\ket{0,0,0,\psi_k}$.
Any eigenstate will work, so the simulation can be initialized, e.g., by making the
wave-function collapse by means of an energy measure.

{\it Step 1:}
A QPE procedure $\Phi^{\rm (old)}$ is performed between register 1 and register 2:
\begin{equation}
  \ket{0}_2 \otimes \ket{\psi_k}_1 \xrightarrow{\Phi^{\rm (old)}} \ket{E_k}_2 \otimes \ket{\psi_k}_1\, ,
\end{equation}
$E_k$ being the energy of the eigenstate $\ket{\psi_k}$.  Thus, the global
output state after this step is $\ket{0,0,E_k,\psi_k}$.

{\it Step 2:}
The choice of the trial update is implemented by randomly selecting a unitary
operator $C$ from a certain set $\mathcal{C}\subseteq \mathcal{U}(2^n)$, and
applying it to the first register\footnote{There is a lot of arbitrariness in
the choice of the set $\mathcal{C}$, but with two requirements: for each
element of $\mathcal{C}$ its inverse must be in $\mathcal{C}$ as well, and the set
must be large enough to ensure ergodicity in the sampling of energy eigenstates.},
\begin{equation}
  \ket{E_k}_2 \otimes \ket{\psi_k}_1 \xrightarrow{C} \sum\limits_p x_{k,p}^{(C)} \ket{E_k}_2 \otimes \ket{\psi_p}_1\, ,
\end{equation}
where $x_{k,p}^{(C)}$ are the matrix elements of $C$.  This operation is
followed by a QPE $\Phi^{\rm (new)}$ between register 1 and register 3:
\begin{equation}
  \ket{0}_3 \otimes \ket{\psi_p}_1 \xrightarrow{\Phi^{\rm (new)}} \ket{E_p}_3 \otimes \ket{\psi_p}_1\, .
\end{equation}
In summary, this brings the output state of step 1 into a new (formal)
superposition of eigenstates
\begin{equation}
    \ket{0,0,E_k,\psi_k} \xrightarrow{\rm Step ~2} \sum\limits_p x_{k,p}^{(C)} \ket{0,E_p,E_k,\psi_p}\, .
\end{equation}

{\it Step 3:}
This step is the quantum analogue of the accept-reject procedure: a one-qubit
operator $W(E_p,E_k)$ is applied to the acceptance register 4, conditioned (by
controlled gates) on both the old and new energy registers (2 and 3), such that 
\begin{equation}
\begin{aligned}
  & \ket{0,E_p,E_k,\psi_p} \xrightarrow{W}  \\
  & \quad \left( \sqrt{f(p,k)}\ket{1} + \sqrt{1-f(p,k)} \ket{0} \right)\otimes \ket{E_p,E_k,\psi_p}\, ,
\end{aligned}
\end{equation}
where
\begin{equation}
  f(p,k) = \min\left(1, e^{-\beta (E_p - E_k)}\right).
\end{equation}
Now, a classical measure is done on the acceptance register, which can only
take the values $1$ or $0$.  If the output of the measure is $1$, one can proceed by
performing a classical measure of energy on register 3, making the system
qubits collapse onto a given energy eigenstate $\ket{1,E_p,E_k,\psi_p}$.
From this state, one resets
\footnote{Resetting a qubit can be implemented, for
example, by taking a classical measure on that qubit, followed by a
$\sigma_x$-gate if the measure returns $1$ (i.e., a CNOT controlled by a classical bit).}
all the non-system qubits (registers 2, 3 and 4) to zero
and then starts again the algorithm from step 1. Things are slightly more complex if the readout of
the acceptance register is $0$: in this case the trial state has to be rejected,
and therefore the update step needs to be reverted, using the procedure
described in step 4.

{\it Step 4:}
First apply the inverse of the operator 
\begin{equation}
U \equiv W \cdot \Phi^{\rm (new)} \cdot C\ ,
\end{equation}
which represents the sequence of unitary operators applied in steps 2 and 3.
Notice that register 3 will be set to zero by $U^\dagger$, while register 2 is
kept untouched from step 1, giving us a reference value for the initial energy.
A QPE $\Phi^{\rm (new)}$ followed by a classical measure on register 3 can
return two possible results: if $E^{\rm new}=E_k$, one has succeeded in
reverting back the system state into an energy eigenstate with
energy\footnote{The system state does not have to be equal to $\ket{\psi_k}$;
indeed, in case of degeneracies, invertibility and ergodicity of the set of
updates $\mathcal{C}$ anyhow guarantee the exactness of the algorithm.
In particular, going to a different state can be viewed as reverting
to the original state followed by a microcanonical step.} 
$E_k$;
it is thus possible to reset all non-system qubits and start again from step 1.
If instead $E^{\rm new} \neq E_k$, one has to apply $[\Phi^{\rm
(new)}]^{-1}$, followed by $U$ and by a measure in the acceptance register (in
order to reach a configuration analogous to the one at the end of step 3), and
repeat step 4 until $E^{\rm new}$ is equal to $E_k$.  From a practical point
of view it is reasonable to set a maximum number of iterations for this
procedure, after which the whole update is aborted and one has to restart from
step 0.

We finally emphasize that the possibility of using quantum gates controlled 
by classical measure is a key ingredient for the implementation of the 
QMS algorithm.
This feature is used at the end of step 3 when, depending on the result of the classical measure on the
acceptance register, one accepts (go to step 1) or rejects (go to step 4) the proposed configuration.
Moreover, in case of reject, another classical control is needed to check whether $E^{\rm new}$ is 
equal to $E^{\rm old}$ or not.

\subsection{Measuring observables}

Using the QMS algorithm, we can generate a sequence of energy eigenstates whose
energy is distributed according to the Boltzmann weight and, in particular, we
can easily compute thermal averages of operators commuting with the Hamiltonian $H$. 
One is also interested, however, in computing thermal averages of
a generic observable $O$, which does not commute with $H$. The
measure of such an observable on the system register would however spoil the
stochastic exactness of the QMS algorithm, since the state register would
collapse away from the given energy eigenstate. 

This problem is obviously absent in the classical algorithm; a
possible way out is to measure the observable $O$ every $n_{th}$
steps of the QMS algorithm, with $n_{th}$ large enough to let the system
rethermalize between consecutive measures. A different strategy could
be to start from step 0 of the QMS algorithm, perform $n_{th}$ updates to
reach thermalization, measure the observable $O$ and finally abort the simulation,
repeating this procedure several times. In this way, we have however less or no
variability in the initial configuration at step 0 (depending on the
initialization procedure adopted), and by performing $n_{th}$ rethermalization
updates after every measure we expect to have a more efficient sampling of the
energies. 

Note that, in both cases, (i.e.~using rethermalization or different
independent simulations) we have to fix the number $n_{th}$, and this
introduces a systematic error in the algorithm. By selecting $n_{th}$ 
large enough, this systematic can be made negligible with
respect to the statistical errors that are present in every MC simulation,
however this means that we cannot improve arbitrarily the accuracy by simply
increasing the statistics.  A detailed discussion of this point will be
presented in Sec.~\ref{systematics}. Finally, let us explicitly remark that, if
we are interested in computing the quantum thermal average of several
operators, rethermalization updates are required after the measure of
each of these observables, the only exception being the case of 
observables commuting with each other, which can be measured (in any order)
with no need for in-between rethermalization.

\section{Numerical Results}
\label{results}

\begin{figure*}
  \includegraphics[width=0.8\linewidth, clip]{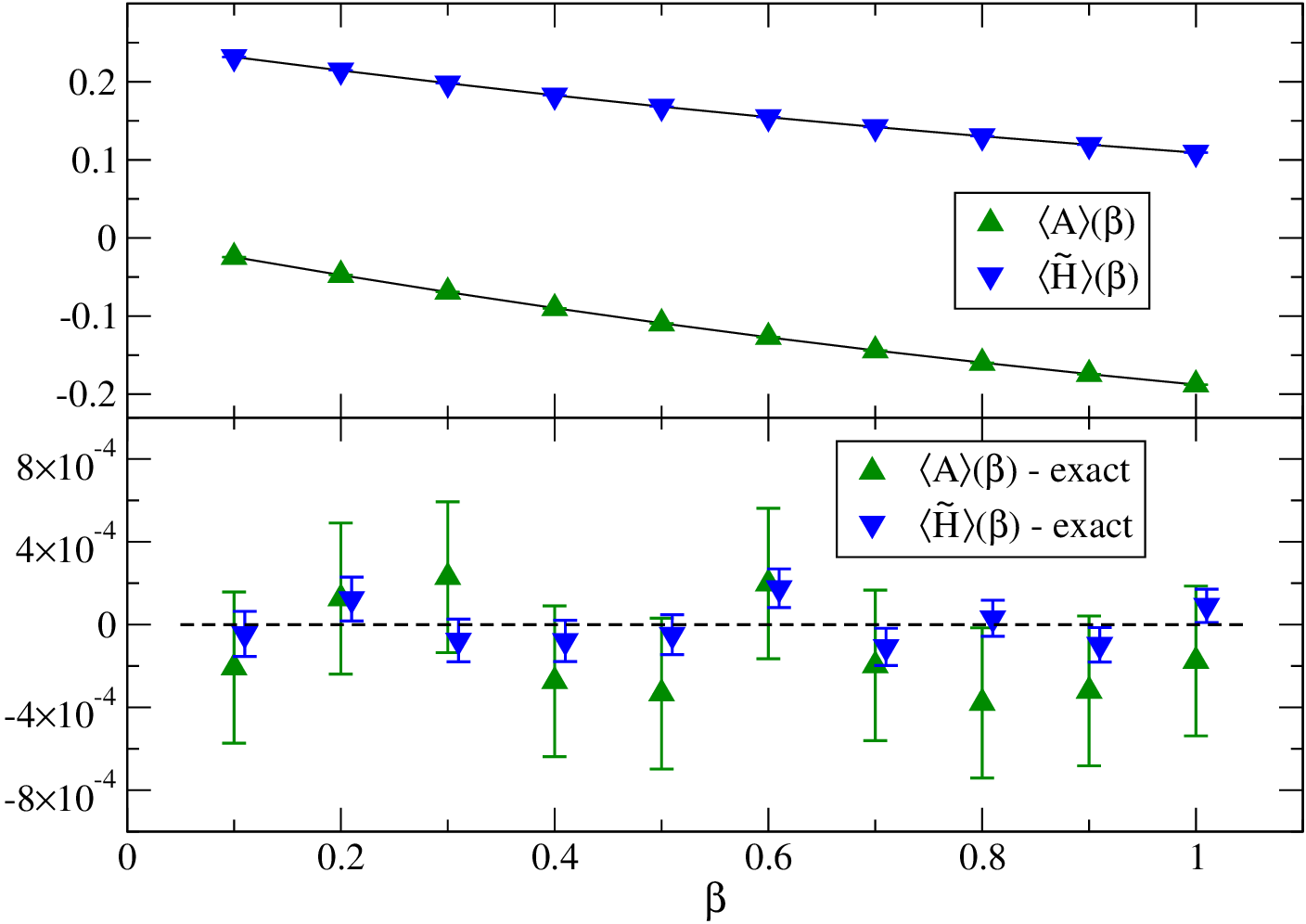}
  \caption{Expectation values of the energy $\tilde H$ and the observable $A$, defined in Eq.~\eqref{eq:A_obs}, for several values of $\beta$. In the top panel, solid lines represent the exact analytical values of the observables; the bottom panel shows residuals (data are slightly shifted horizontally, to improve readability), in order to better appreciate the consistency
between analytical and numerical results.}
  \label{fig:EA_plot}
\end{figure*}

In this section we present the results obtained by applying the QMS algorithm
to the frustrated triangle system described by the Hamiltonian in
Eq.~\eqref{eq:HamSpinSys}.

States of the frustrated triangle can be represented using $n=3$ qubits, while
the energy of the system can be exactly encoded by using only one qubit
($r=1$). Indeed, since the energy takes only two values ($-J$ and $3J$), by
shifting and rescaling $H$ we can define the new Hamiltonian
$\tilde H = (\Id + H/J)/4$, whose spectrum is only made of eigenvalues
equal to $0$ or $1$. The total number of required qubits is therefore $n+2r+1=6$.

To proceed with the QMS algorithm we need two basic ingredients: to be able to
perform a QPE and to determine the set $\mathcal{C}$ of unitary operators to be
used for generating trial states. As previously discussed, the set
$\mathcal{C}$ has to ensure ergodicity (in this context, by ergodicity 
we do not mean the possibility to actually explore the whole Hilbert
space, but only the whole set of energy eigenstates)
and if $U\in \mathcal{C}$, also $U^{\dag}$ has to belong to the same set (reversibility).

The naive choice of single spin flips $\Id\otimes\Id\otimes\sigma_x$,
$\Id\otimes\sigma_x\otimes\Id$ and $\sigma_x\otimes\Id\otimes\Id$ does not
ensure ergodicity, as evident from the observation that all these operators
commute with $H$. An ergodic choice can be obtained instead by replacing the
spin flip operator $\sigma_x$ in the previous expressions by the Hadamard gate
$\frac{1}{\sqrt{2}}(\sigma_z+\sigma_x)$  \cite{NielsenChuang}. Since these
transformations are involutions, reversibility is also automatically satisfied. 
With this choice, we obtain an average acceptance probability in the
Metropolis step that goes from $99\%$ for $\beta = 0.1$ to 
$91\%$ for $\beta = 1.0$.

For what concerns the QPE, this actually boils down to a real-time evolution
(see Sec.~\ref{sec:QPE}) and, for the frustrated triangle, time evolution is
particularly simple: since all terms in $H$ commute pairwise and their
square is proportional to the identity, we can easily construct the time
evolution operator as a product of terms like
\begin{align}
  e^{-i t J \Id \otimes \sigma_x \otimes \sigma_x } & = \big( \Id \otimes \Id \otimes \Id \big) \cos(J t) \nonumber \\
  & -i \big( \Id \otimes \sigma_x \otimes \sigma_x \big) \sin(J t)\ .
\end{align}
We will come back later to the systematics associated to the QPE when such an
analytical expression is not available.

Now we have all the ingredients required for the application of the QMS
algorithm and, in order to test its effectiveness, we start by studying the
dependence on the temperature of the internal energy $\langle \tilde H
\rangle$, comparing numerical data with exact analytical results. Such a comparison is
performed in Fig.~\ref{fig:EA_plot} and we can see that there is excellent
agreement between the two determinations. 

We now move to the computation of thermal averages of operators not
commuting with the Hamiltonian. We have tested several Hermitian operators, 
but here we only report the results obtained for
\begin{equation}\label{eq:A_obs}
    A = \sigma_x \otimes \sigma_x \otimes {\big( \Id+ \sigma_y
    \big)}\, .
\end{equation}
The outcomes obtained in all other cases are equivalent. After
some preliminary test (see Sec.~\ref{sec:Syst-Retherm} for a thorough
analysis) we decided to use 200 rethermalization steps after each measure of
$A$, and in Fig.~\ref{fig:EA_plot} we show the results obtained with a
statistics of about $10^7$ measures. Even in this case, the agreement between
the numerical estimates and the exact results is remarkable;
this is a direct proof of the fact that the algorithm properly samples the configurations
and effectively solves the sign problem of this model. 

\begin{figure}[b]
  \includegraphics[width=1.0\linewidth, clip]{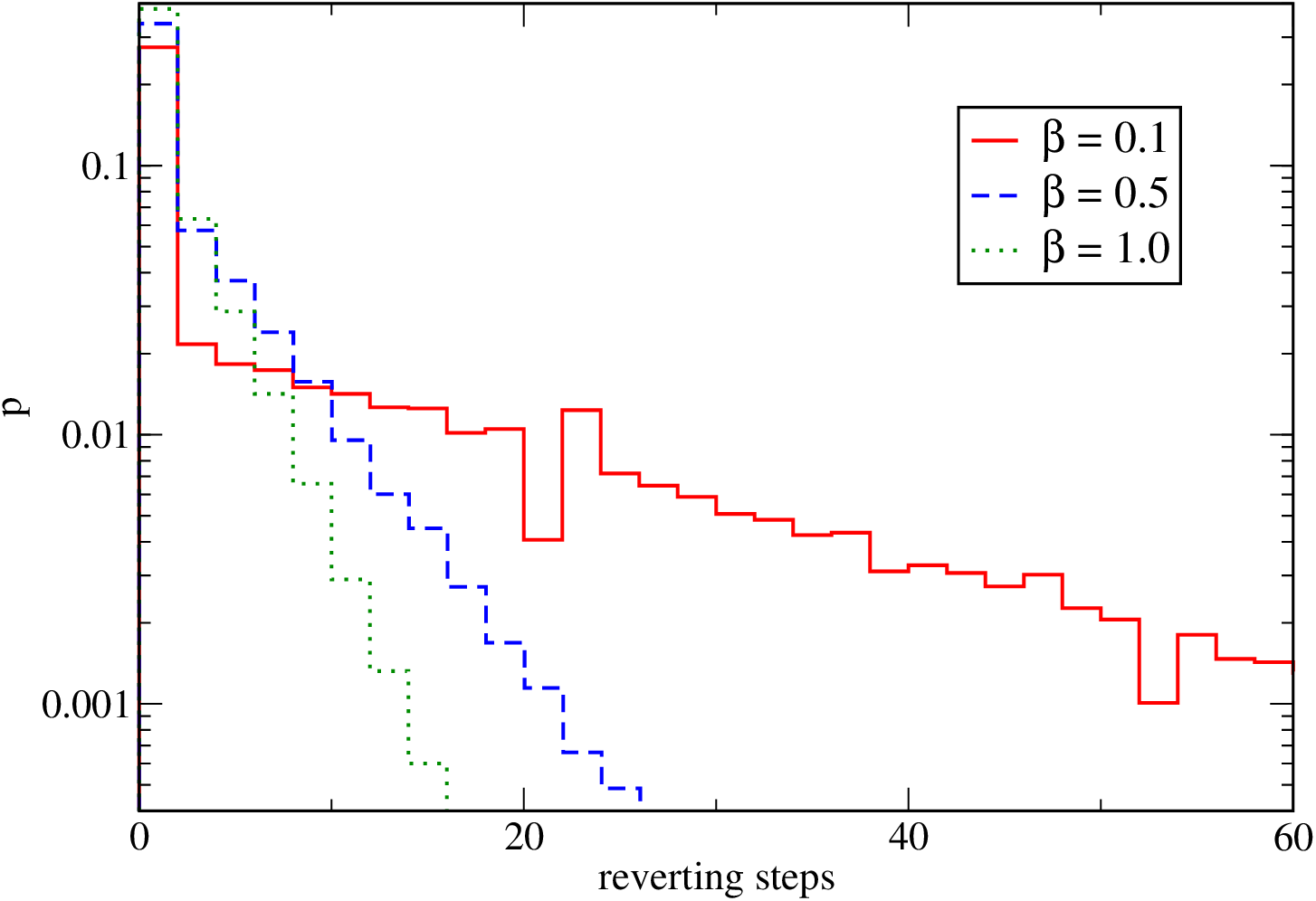}
  \caption{Histogram of the number of times step 4 of the QMS algorithm had to be repeated, for 
    $\beta = 0.1, 0.5$ and $1.0$.}%
  \label{fig:revsteps}
\end{figure}

After having checked that the QMS algorithm correctly samples the states of the
system, we can investigate some aspects related to the efficiency of the
method. In particular, a possible weak point of the QMS algorithm is the
iterative procedure required in the step 4 (see Sec.~\ref{sec:QMS}) to project
back a rejected update: if a very large number of iterations is required, the
algorithm obviously loses its effectiveness. However we found that, at least
for the frustrated triangle studied in this work, the number of reverting steps
required for each iteration is relatively small, and its distribution can be
well approximated by an exponential, as shown in the histogram of
Fig.~\ref{fig:revsteps} for some values of $\beta$. The distribution of the
number of reverting steps gets broader for small $\beta$ values, signaling that
the algorithm is less efficient in this regime; on the other hand, this is
mitigated by the fact that at low $\beta$ the acceptance is higher, so that
reverting steps occur less frequently.  A possible way of reducing the impact
of the reverting steps on the simulation time would be to look for a set
$\mathcal{C}$ of moves with larger acceptance probability, so to reduce the
occurrence of the iterative step.

One last comment must be done on the number of quantum gates required
to perform a single Metropolis step. In order to perform the QPE 
and apply the operator $U$, i.e.\ before measuring
the acceptance register, we need $\sim\!\!100$ gates. If we accept the new configuration we need 
additional $\sim\!\!200$ gates to go back to step 1, otherwise, in case of rejection,
we need $\sim\!\!100$ gates in order to perform a single reverse attempt.
In the counting of the quantum gates all the classical measures and all the quantum gates controlled by 
a classical measure have been excluded. We see that, even for a very simple system like the frustrated 
triangle, the number of quantum gates used is large, therefore, in order to
run the algorithm on a real quantum computer, a very high fidelity on the quantum gates is required.

\section{Systematics of the QMS algorithm}
\label{systematics}

The QMS algorithm generally presents a certain number of
systematic errors. In the previous section, we already discussed the
rethermalization updates to be performed after each measure of an observable
not commuting with the Hamiltonian. For the frustrated triangle this was the
only source of systematic error in the QMS algorithm. Here we discuss the
systematics that have to be taken into account, beyond rethermalization,
when using QMS to simulate a generic quantum system.

\emph{Digitalization of the energy}: in the QMS update, energy has to be stored in two
quantum registers, each composed of $r$ qubits, and if the spectrum of the
Hamiltonian cannot be exactly mapped to an $r-$bit binary number, one
introduces a truncation error in the energies and thus a systematic in the
sampling. Even for toy models, this is clearly unavoidable whenever energy
eigenvalues are incommensurable with each other, like for a three-state
system with energy levels $0, 1/\sqrt{2}$, and $1$.

\emph{Digitalization of the state}: this problem is present for systems with
continuous degrees of freedom (like gauge theories associated to continuous Lie
groups). In such cases we need to choose a proper digitalization of the state,
i.e. a way of approximating the state by using only $n$ qubits. 
For more details, 
see Refs.~\cite{Klco:2018zqz,Alexandru:2019ozf,Alexandru:2019nsa,Klco:2019evd}.

\emph{QPE implementation}: as briefly recalled in Sec.~\ref{sec:QPE}, the
implementation of the QPE algorithm requires a real-time unitary evolution on the system
register using $e^{-i H t}$. In practice, such an evolution cannot be performed
exactly, but has to be approximated by a Trotterization, thus introducing
a systematic error related to the finite time-step (see, e.g.,
Refs.~\cite{Berry2007,Childs2018,Childs2019,Lamm:2019bik,Chakraborty:2020uhf,Kharzeev:2020kgc}
for recent discussions about this systematic effect in similar contexts).

The effect of all these systematics can be reduced at the expense of increasing
the computational complexity of the algorithm: by increasing the number of
used qubits, the digitalization of both energies and states can be made
arbitrarily accurate and similarly the Trotter time-step can be reduced at will,
making the QPE more and more accurate. While this is true in theory, in
practice one is interested in understanding how these systematics scale with
the computational complexity, in order to understand whether the algorithm is useful
or not for the problem at hand. 

The last two systematics of the list (i.e. digitalization of the state and QPE
implementation) are not specific of the QMS, but are common to all QC
algorithms. For this reason, they have already been discussed in the literature
and we will not further investigate them, thus concentrating
on the rethermalization and the digitalization of energy. 

\begin{figure}[t]
    \centering
    \includegraphics[width=1.0\linewidth, clip]{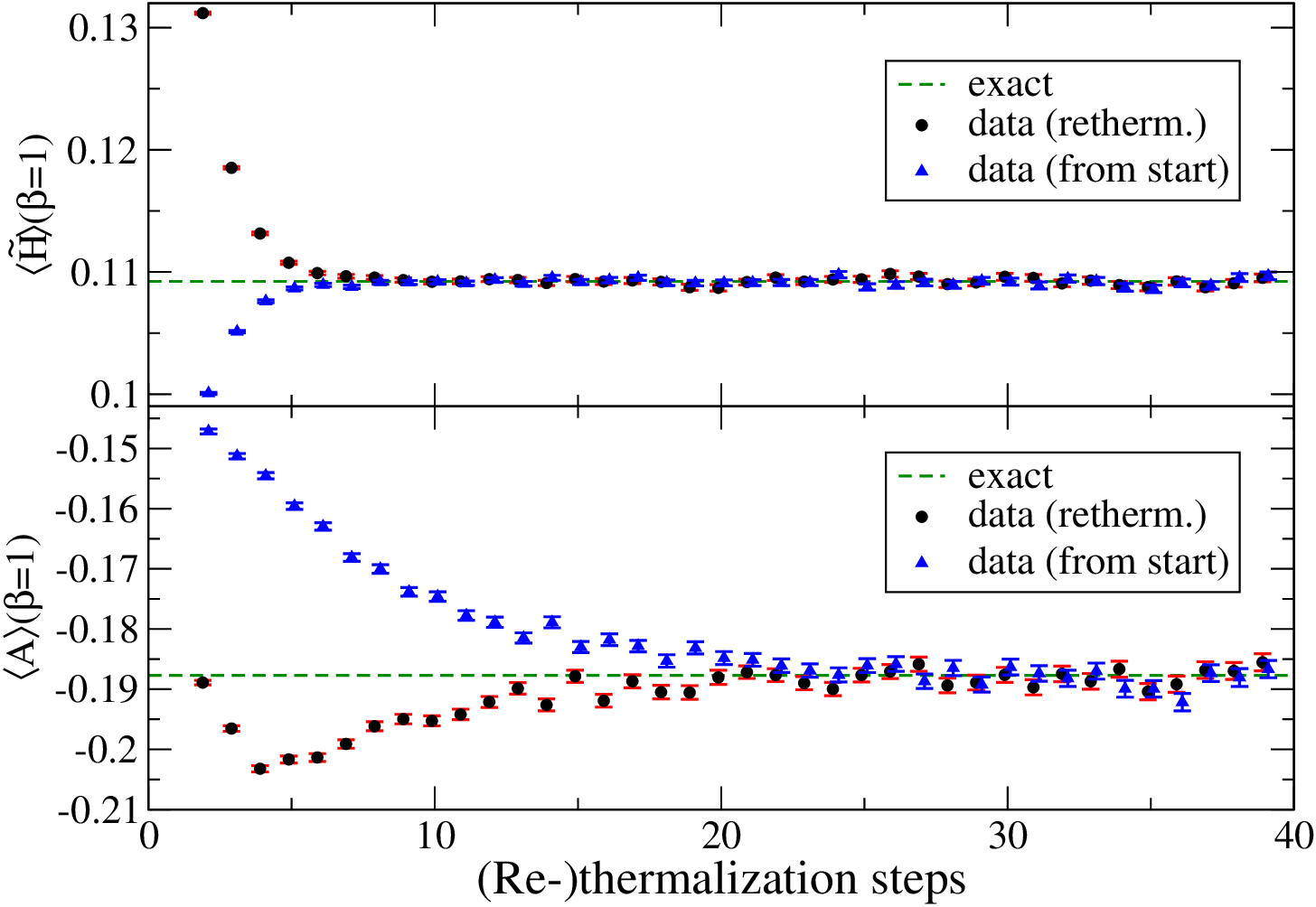}
    \caption{Effects of different numbers of 
      (re-)thermalization steps at $\beta = 1$ on the expectation values
      of the energy and of the non-$H$-commuting observable $A$ defined in Eq.~\eqref{eq:A_obs}.
      The points with a circle refer to measures 
      taken during the Metropolis chain after each rethermalization sequence,
      while the ones with a triangle refer to measures taken at the end of a single Metropolis chain
      after a certain number of thermalization steps, starting always
      from the initial eigenstate $\frac{1}{\sqrt{2}} (\ket{\mathbf{0}}-\ket{\mathbf{3}})$.}
    \label{fig:avestd_EA_b1_compare}
\end{figure}

\subsection{Rethermalization}
\label{sec:Syst-Retherm}
We have performed several simulations with different numbers of
rethermalization steps, to quantify the number of needed rethermalization
updates.  As an example, the results obtained for the average values $\langle
\tilde{H}\rangle$ and $\langle A\rangle$ [defined in Eq.~\eqref{eq:A_obs}] are shown in
Fig.~\ref{fig:avestd_EA_b1_compare} for $\beta = 1$.  The observed behavior is
analogous to that seen in the thermalization of classical MC simulations:
average values depend on the number of rethermalization steps $n_{th}$ adopted,
but they converge to an asymptotic value when $n_{th}$ increases. 

This asymptotic value is the correct thermal average, which can however be
reached at different values of $n_{th}$ for different observables, as clearly
seen in Fig.~\ref{fig:avestd_EA_b1_compare}; this is just a consequence of the
fact that different observables have different thermalization times. In order
not to introduce systematics in the final estimates, $n_{th}$ has to be chosen
large enough that all the observables have reached their plateau values with an
accuracy that is larger than the target precision of the simulation.

\subsection{Digitalization of the energy}
\label{sec:Syst-E}

In order to assess how much the QMS algorithm relies on an accurate estimation
of the energies, we considered a simple system made up of two qubits representing
an Hamiltonian $H$ with the following four energy levels:
$\{0, \frac{1}{2}, \frac{1}{\sqrt{2}}, \frac{3}{4}\}$.
With this setup, energy levels can never be exactly digitalized, and we can also
investigate the effect of having two nearby levels ($1/\sqrt{2}$ and $3/4$)
that cannot be resolved if the number of used qubits is too small. 

\begin{figure}[!t]
  \includegraphics[width=1.0\linewidth, clip]{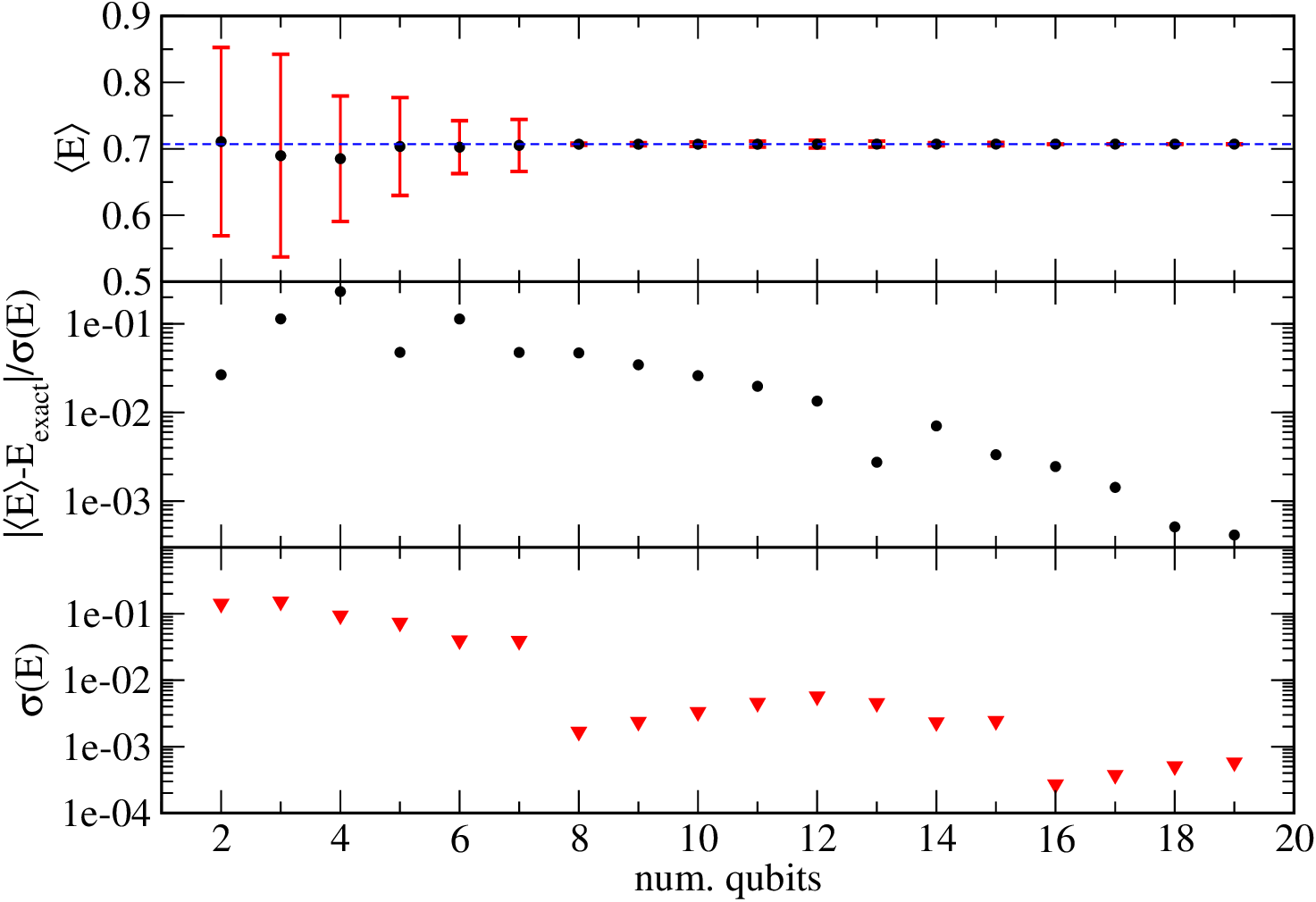}
  \caption{Accuracy of the QPE procedure for the exact eigenvalue 
    $\lambda_1 = \frac{1}{\sqrt{2}}$ using different numbers of qubits for
    the energy register.
    Here $\langle E \rangle$ and $\sigma(E)$ denote the average value and the
    standard deviation of the energy distribution, as obtained from QPE.
    In the upper panel, QPE estimates for $\langle E \rangle$ are compared with
    the exact value (dashed line); the middle panel shows the ratio between the 
    absolute value of the discrepancy and the standard deviation;
    the bottom panel shows the scaling of $\sigma(E)$ with the number of qubits.}
  \label{fig:PE_accuracy_isolated}
\end{figure}

As a preliminary test, we isolated the QPE step: in
Fig.~\ref{fig:PE_accuracy_isolated} we show the results of a single
application of the QPE procedure $\Phi$ to the eigenstate of $H$ associated to
the eigenvalue $\lambda_1 = \frac{1}{\sqrt{2}}$, using a variable number $r$ of
qubits for the energy register.  Since $\lambda_1$ cannot be represented by a
fraction of powers of $2$, a measure in the energy register does not always 
collapse
to the same state, but has a certain probability to collapse to a state
with neighboring energy, so that each measure is affected by 
a sampling error, which however vanishes exponentially in $r$.

\begin{figure}[t]
    \centering
    \includegraphics[width=1.0\linewidth, clip]{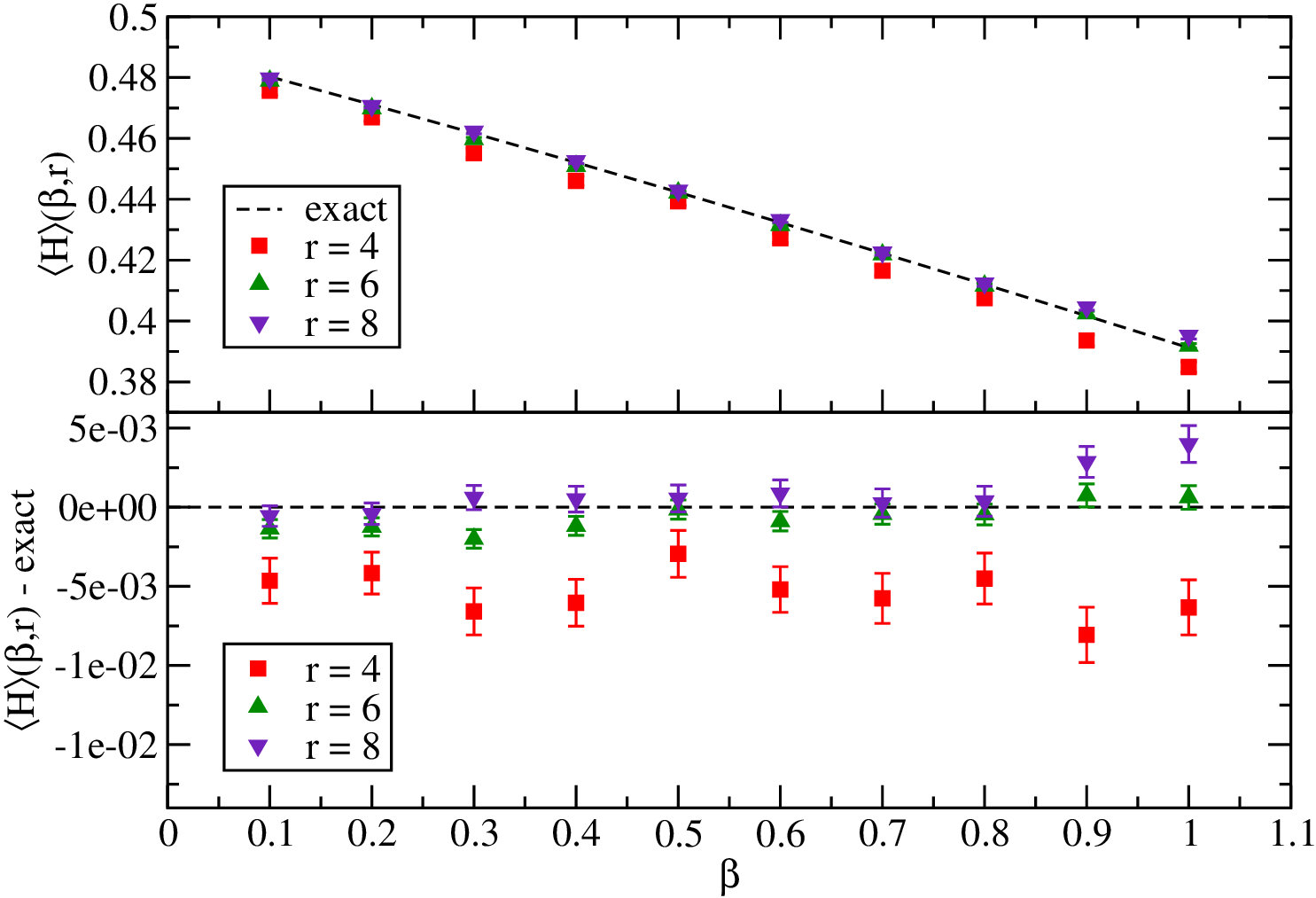}
    \caption{Energy sampling for a system with energy levels
      $\{0, \frac{1}{2},  \frac{1}{\sqrt{2}}, \frac{3}{4}\}$, using a different
      number of qubits for the energy registers ($r=4,6$ and $8$), and comparison
      with the exact expectation value $\langle H(\beta)\rangle$
      for several values of $\beta$.}
    \label{fig:PE_means}
\end{figure}

\begin{figure}[!t]
  \includegraphics[width=1.0\linewidth, clip]{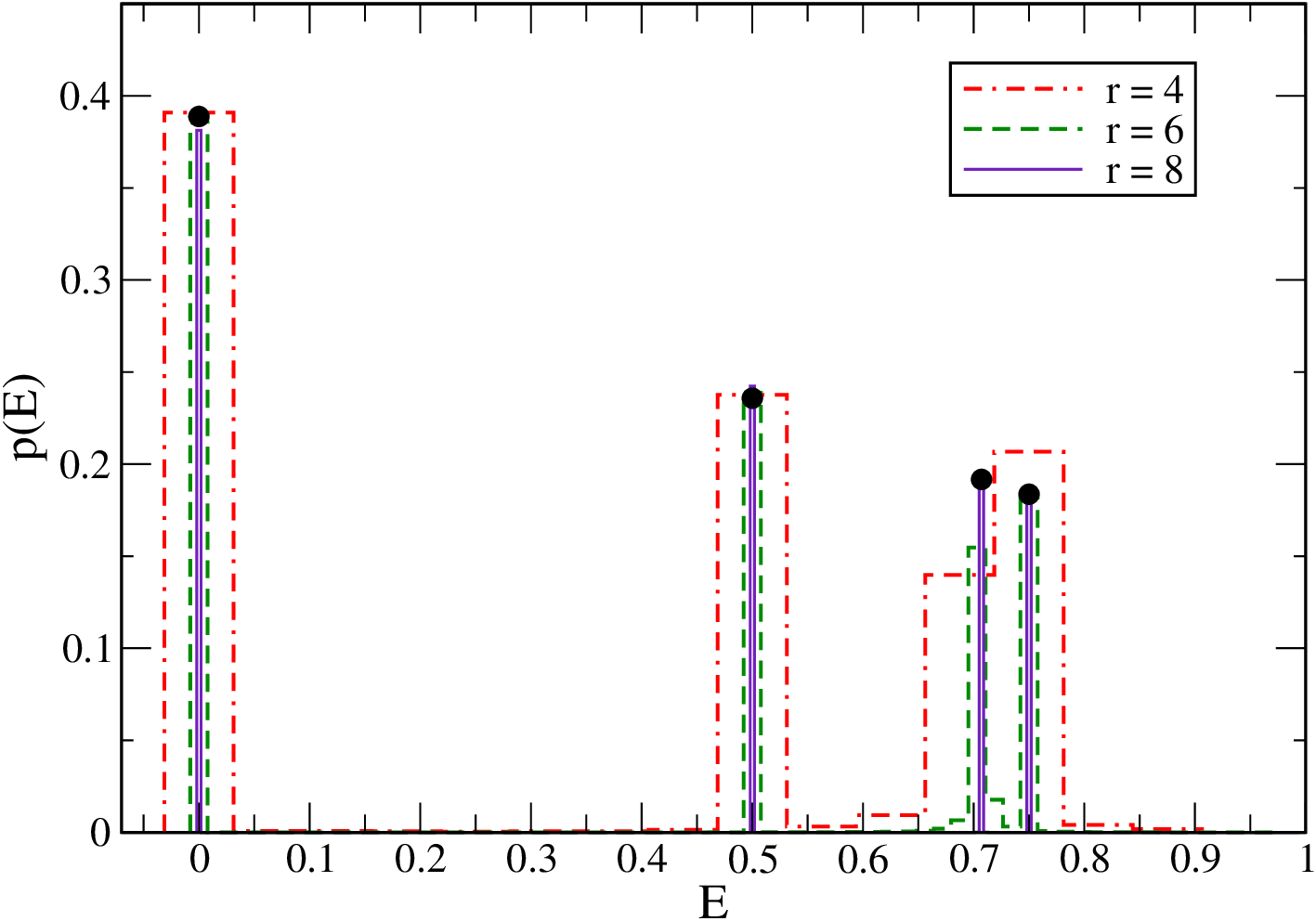}
  \caption{Energy sampling at $\beta=1.0$ with energy levels
    $\{0, \frac{1}{2}, \frac{1}{\sqrt{2}}, \frac{3}{4}\}$, using different
    numbers of qubits for the energy registers ($r=4,6$ and $8$),
    and comparison with the exact probability distribution $p(E)$.
  For each choice of $r$, around $O(10^4)$ measures have been performed.}%
  \label{fig:Eprob}
\end{figure}

More interesting are the effects of errors in the estimates occurring during 
the full QMS algorithm, as can be observed from {the average of
the energies measured, shown in Fig.~\ref{fig:PE_means} for three different digitalizations
of the energy ($r=4,6$, and $8$).
Using $4$ qubits seems not to give a sufficiently accurate estimate along the explored range
in $\beta$, while $6$ and $8$ qubits show reasonable accuracy, even if
the convergence behavior is not always clear, as apparent, for example, at $\beta=1$.
Nevertheless, comparing the energy probability distribution of the samples with the exact
one (Fig.~\ref{fig:Eprob}), we can observe a clearer convergence to the exact weights
at increasing number of qubits $r$.} 
Notice that a proper discrimination between neighboring eigenvalues in
the spectra (as $\frac{1}{\sqrt{2}}$ and $\frac{3}{4}$) can occur only if the number
of qubits is sufficient.

\section{Conclusions}
\label{conclusions}

In this paper we have described, implemented, and tested a quantum algorithm
able to measure thermal averages for quantum mechanical systems. Our results
are for a very simple system of frustrated spins, that, if rewritten in terms
of Euclidean path integrals exhibits the infamous sign problem, that makes
the use of classical Monte Carlo simulations impossible. In this sense, our toy
model shares the same problems of more complex and physically meaningful
problems, such as the study of the phase structure of Lattice QCD at finite
temperature and density. It therefore conceptually paves the way to quantum
simulations of those systems, as soon as appropriately powerful quantum
computers become available in the future.

The algorithm chosen for our study is the 
Quantum Metropolis Algorithm introduced in Ref.~\cite{QMS_paper}, which 
has been briefly reviewed in Sec.~\ref{qmsalgo}. 
Even if inspired by the standard Metropolis algorithm, it cannot 
be considered a Markov chain in the classical sense, since the 
measurement of observables interferes with the Markov process itself,
thus destroying equilibration and leading to the need for rethermalization
after each measurement is taken (unless the observable commutes with the 
Hamiltonian). Of course this is not a peculiar feature of the algorithm,
but rather something which one expects in general, because of the 
quantum nature of the system and of the algorithm.

We have verified that our quantum algorithm reproduces, 
within statistical errors,
exact results for the thermal averages of the energy and of an additional
operator that does not commute with the Hamiltonian. We have quantified the
size of the statistical and sampling errors and their behavior with respect to the
number of simulation steps. We have also discussed and measured the systematic
errors due to several different effects, such as the accuracy of the phase
estimation process, the sampling accuracy as a function of the qubit
register-size, as well as some conceptually unavoidable performance losses
associated to the reverting steps of the algorithm.

All these systematic errors, together with the computing depth of the complete
algorithm, will ultimately play a key role to define the actual viability of
this quantum algorithm as one moves from simple toy models to the study of 
real-life
complex physical systems. This perspective suggests several avenues for further
studies, ranging from the implementation of the present model onto actual QC
machines, to the implementation of the algorithm for more complex (possibly
gauge invariant) models, to the analysis of the optimal balance point between
computational accuracy and efficiency versus register-size and algorithm
complexity: work is in progress along these directions.

\acknowledgments

We thank Oliver Morsch and Ivano Tavernelli for useful discussions.
Numerical simulations have been performed at the 
Scientific Computing Center at INFN-PISA.
DV acknowledges support from the INFN HPC-HTC project.

\appendix

\section{Derivation of Eq.~\eqref{eq:rewrite}}
\label{app:A}

Here we present a derivation of Eq~(\ref{eq:rewrite}), 
which was used in Sec.~\ref{description} to better understand
the structure of the path-integral representation of the frustrated triangle.
We start from the Taylor expansion of the Euclidean evolution operator:
\begin{equation}
  \mathrm{e}^{-\frac{\beta H}{N}} = \sum
  _{n=0}^{\infty}\frac{1}{n!}{\left(-\frac{\beta}{N}\right)}^nH^{n}\ .
  \label{eq:taylor_exp}
\end{equation}
It is then easy to check that the Hamiltonian~\eqref{eq:HamSpinSys} satisfies the property
$H^2 = (3J^2\Id + 2JH)$, therefore we have, for any integer $n$, a relation of the form 
\begin{equation}
  H^n = \alpha _n\Id + \beta _{n} H\ .
\end{equation}
To fix the values of $\alpha_n$ and $\beta_n$ we can diagonalize both 
members of this equation and remember that the eigenvaules of $H$ are just $-J$
and $3J$.  In this way, we immediately get
\begin{equation}
  H^n = \frac{J^n}{4} \left\{ \big[ 3(-1)^n + 3^n \big] \Id + \big[ 3^n + (-1)^{n+1} \big] \frac{H}{J} \right\}
\end{equation}
and, plugging this result into Eq.~(\ref{eq:taylor_exp}) we finally
obtain 
\begin{eqnarray}
  \mathrm{e}^{-\frac{\beta H}{N}} &=& 
  \sum_{n=0}^{\infty}\frac{1}{n!}\left(-\frac{\beta}{N}\right) ^n H^{n} \\
  &=& \frac{1}{4}\left[\left(e^{-3\frac{\beta J}{N}} + 3e^{\frac{\beta J}{N}}
    \right)\Id
    + \left(e^{-3\frac{\beta J}{N}} - e^{\frac{\beta J}{N}} \right)\frac{H}{J} \right] \ . \nonumber
\end{eqnarray}

\end{document}